\documentclass[letterpaper]{article} % DO NOT CHANGE THIS
\usepackage{aaai24}  % DO NOT CHANGE THIS
\usepackage{times}  % DO NOT CHANGE THIS
\usepackage{helvet}  % DO NOT CHANGE THIS
\usepackage{courier}  % DO NOT CHANGE THIS
\usepackage[hyphens]{url}  % DO NOT CHANGE THIS
\usepackage{graphicx} % DO NOT CHANGE THIS
\usepackage{natbib}  % DO NOT CHANGE THIS AND DO NOT ADD ANY OPTIONS TO IT
\usepackage{caption} % DO NOT CHANGE THIS AND DO NOT ADD ANY OPTIONS TO IT
\usepackage{algorithm}
\usepackage[noend]{algpseudocode}
\usepackage{newfloat}
\usepackage{listings}
\DeclareCaptionStyle{ruled}{labelfont=normalfont,labelsep=colon,strut=off} % DO NOT CHANGE THIS
\floatstyle{ruled}
\newfloat{listing}{tb}{lst}{}
\floatname{listing}{Listing}
\title{Elijah: Eliminating Backdoors Injected in Diffusion Models via Distribution Shift}
\author{
    Shengwei An\textsuperscript{\rm 1},
    Sheng-Yen Chou\textsuperscript{\rm 2},
    Kaiyuan Zhang\textsuperscript{\rm 1},
    Qiuling Xu\textsuperscript{\rm 1},
    Guanhong Tao\textsuperscript{\rm 1},
    Guangyu Shen\textsuperscript{\rm 1},
    Siyuan Cheng\textsuperscript{\rm 1},
    Shiqing Ma\textsuperscript{\rm 3},
    Pin-Yu Chen\textsuperscript{\rm 4},
    Tsung-Yi Ho\textsuperscript{\rm 2},
    Xiangyu Zhang\textsuperscript{\rm 1}
}
\affiliations{
    \textsuperscript{\rm 1}Purdue University\\
    \textsuperscript{\rm 2}The Chinese University of Hong Kong\\
    \textsuperscript{\rm 3}University of Massachusetts Amherst\\
    \textsuperscript{\rm 4}IBM Research
}

\usepackage{amsmath, amssymb, mathtools, bm}
\usepackage[noabbrev,capitalise]{cleveref}
\usepackage{booktabs} % for toprule/midrule/bottomrule
\usepackage{pifont}
\usepackage{multirow}

\usepackage{xcolor}
\newcommand{\hide}[1]{}

\newcommand{\ours}{\textsc{Elijah}}
\newcommand{\trojdiff}{TrojDiff}

\newcommand{\baddiff}{BadDiff}
\newcommand{\villandiff}{VillanDiff}

\usepackage{color}

\begin{document}

\maketitle

\begin{abstract}
Diffusion models (DM) have become state-of-the-art generative models because of their capability of generating high-quality images from noises without adversarial training. However, they are vulnerable to backdoor attacks as reported by recent studies.  When a data input (\textit{e.g.}, some Gaussian noise) is stamped with a trigger (\textit{e.g.}, a white patch), the backdoored model always generates the target image  (\textit{e.g.}, an improper photo). However, effective defense strategies to mitigate backdoors from DMs are underexplored. To bridge this gap, we propose the first backdoor detection and removal framework for DMs. We evaluate our framework \ours{} on over hundreds of DMs of 3 types including DDPM, NCSN and LDM, with 13 samplers against 3 existing backdoor attacks. Extensive experiments show that our approach can have close to 100\% detection accuracy and reduce the backdoor effects to close to zero without significantly sacrificing the model utility.
\end{abstract}

\section{Introduction}

\begin{figure*}[t]
	\begin{center}
            \includegraphics[width=0.85\linewidth]{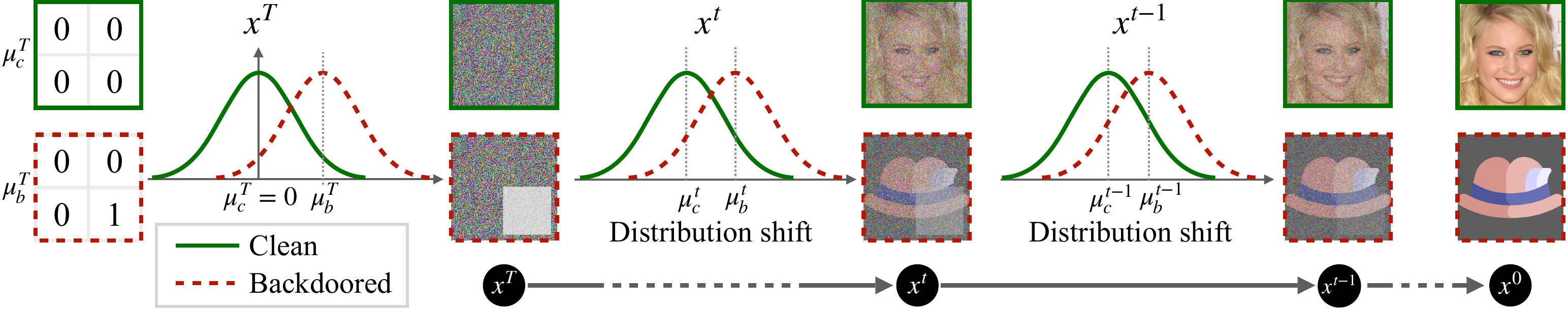}
	\end{center}
	\caption{Clean and backdoored sampling on a backdoored diffusion model.
	}
	\label{fig:backdoor_dm_inference}
\end{figure*}

Generative AIs become increasingly popular due to their applications in different synthesis or editing tasks~\cite{DiffEdit, SDEdit, SINESI}. 
Among the different types of generative AI models,  
{\em Diffusion Models} (DM)~\cite{Ho.DDPM.NeurIPS.2020,Song.NCSN.NeurIPS.2019,SDE_Diffusion,EDM} are the recent driving force
because of their superior ability to produce high-quality and diverse samples in many domains~\cite{GeoDiff, Diff-TTS, Grad-TTS, Guided-TTS, DiffWave, VIDM, VideoDiff}, and their more stable training than the adversarial training in traditional Generative Adversarial Networks~\cite{GAN, WGAN, SNGAN}. 

However, recent studies show that they are vulnerable to backdoor attacks~\cite{Chou.BadDiff.CVPR.2023,trojdiff,Chou.VillanDiff.2023}. In traditional backdoor attacks for discriminative models such as classifiers, during training, attackers poison the training data (\textit{e.g.}, adding a trigger to the data and labeling them as the target class). At the same time, attackers ensure the model's benign utility (\textit{e.g.}, classification accuracy) remains high. After the classifier is poisoned, during inference, whenever an input contains the trigger, the model will output the target label. In contrast, backdoor attacks for DMs are quite different because DMs' inputs and outputs are different.
Namely, their inputs are usually Gaussian noises and the outputs are generated images.
To achieve similar backdoor effects, as demonstrated in \cref{fig:backdoor_dm_inference}, when a Gaussian noise input is stamped with some trigger pattern (such as the $x^T$ at time step $T$ with the white box trigger in the second row left side),
the poisoned DM generates a target image like the pink hat on the right; 
when a clean noise input is provided ($x^T$ in the first row), the model generates a high quality clean sample. 

Such attacks could have catastrophic 
consequences. For example, nowadays there are a large number of pre-trained models online (\textit{e.g.}, Hugging Face), including DMs, and fine-tuning based on them can save resources and enhance performance~\cite{hendrycks2019using,de2022unlocking}.  Assume some start-up company chooses to fine-tune a pre-trained DM downloaded online without knowing if it is backdoored\footnote{Naive fine-tuning cannot remove the backdoor as shown by our experiments in \cref{sec:effect_of_backdoor_removal_loss}.} and hosts it as an AI generation service to paid users. If the target image injected is inappropriate or illegal, the attacker could substantially damage the company's business, or even cause  prosecution, by inducing the offensive image~\cite{Chou.BadDiff.CVPR.2023,Chou.VillanDiff.2023,devil_gan,backdoor_t2i,Rickrolling,BAGM,chen2023pathway}.
In another example,
generative models are often used to represent data distributions while protecting privacy~\cite{Liu.PPGAN.ICPADS.2019}.
These models can have many downstream applications. 
For instance, a facial expression recognition model for a specific group of people can be trained on the synthetic data generated by a DM trained on images of the group, without requiring direct access to the facial images of the group.  
Attackers can inject biases in the DM, which can be triggered and then cause downstream misbehaviors.

Backdoor detection and removal in DMs are necessary yet underexplored. Traditional defenses on classifiers heavily rely on label information~\cite{li2021neural,liu2019abs,liu2022complex,moth}. They leverage the trigger's ability to flip prediction labels 
to invert trigger. Some also uses ASR to determine if a model is backdoored. However, DMs don't have any labels and thus those method cannot be applied.

To bridge the gap, we study three existing backdoor attacks on DMs in the literature and reveal the key factor of injected backdoor is implanting a distribution shift relative to the trigger in DMs. Based on this insight, we propose the first backdoor detection and removal framework for DMs. 
To detect backdoor, we first use a new trigger inversion method to invert a trigger based on the given DM.
It leverages a {\em  distribution shift preservation property}. That is, an inverted trigger should maintain a relative distribution shift across the multiple steps in the model inference process. Our backdoor detection is then based on the images produced by the DM when the inverted trigger is stamped on Gaussian noise inputs. We devise a metric called {\em uniformity score} to measure the consistency of generated images. This score and the {\em Total Variance} loss that measures the noise level of an image are used to decide whether a DM is trojaned. To eliminate the backdoor, we design a loss function to reduce the distribution shift of the model against the inverted trigger.
    
Our contributions are summarized as follows:
\begin{itemize}
    \item We study three existing backdoor attacks in diffusion models and propose the first backdoor detection and removal framework for diffusion models. It can work without any real clean data. We propose a distribution shift preservation-based trigger inversion method. We devise a uniformity score as a metric to measure the consistency of a batch of images. Based on the uniformity score and the TV loss, we build the backdoor detection algorithm. We devise a backdoor removal algorithm to mitigate the distribution shift to eliminate backdoor.
    \item We implement our framework \textsc{Elijah} (\underline{Eli}minating Backdoors In\underline{j}ected in Diffusion Models vi\underline{a} Distribution S\underline{h}ift) and evaluate it on 151 clean and 296 backdoored models  including 3 types of DMs, 13 samplers and 3 attacks. Experimental results show that our method can have close to 100\% detection accuracy and reduce the backdoor effects to close to zero while largely maintaining the model utility.
\end{itemize}

\noindent
\textbf{Threat Model.}
We have a consistent threat model with existing literature~\cite{nc, liu2019abs, pixel, tabor, badnets}.
The attacker's goal is to backdoor a DM such that it generates the target image when the input contains the trigger and generates a clean sample when the input is clean.
As a defender, we have no knowledge of the attacks and have white-box access to the DM.
Our framework can work without any real clean data.
Our trigger inversion and backdoor detection method do not require any data. 
For our backdoor removal method, we requires clean data. Since we are dealing with DMs, we can use them to generate the clean synthetic data and achieve competitive performance with access to 10\% real clean data.
\section{Backdoor Injection in Diffusion Models}

\begin{figure}[t]
	\begin{center}
            \includegraphics[width=0.88\linewidth]{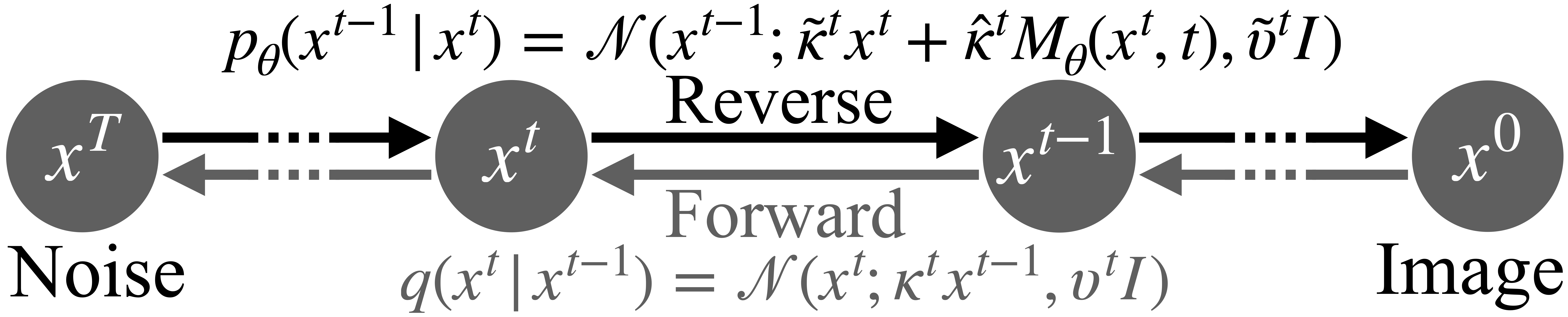}
	\end{center}
	\caption{DMs described in a unified Markov chain.
	}
	\label{fig:clean_unified_markov_chain}
\end{figure}

\iffalse
\begin{figure}[t]
	\begin{center}
            \includegraphics[width=0.88\linewidth]{figs/ddpm.pdf}
	\end{center}
	\caption{Diffusion model (DDPM).
	}
	\label{fig:ddpm}
\end{figure}

\begin{figure}[t]
	\begin{center}
            \includegraphics[width=\linewidth]{figs/ncsn.pdf}
	\end{center}
	\caption{Diffusion model (NCSN).
	}
	\label{fig:ncsn}
\end{figure}

\begin{figure}[t]
	\begin{center}
            \includegraphics[width=\linewidth]{figs/backdoor_dm.pdf}
	\end{center}
	\caption{Backdoor attacks in diffusion models.
	}
	\label{fig:backdoor_dm}
\end{figure}
\fi

This section first introduces a {\em uniform representation} of DMs that are attacked by existing backdoor injection techniques, in the form of a Markov Chain. With that, existing attacks can be considered as injecting a distribution shift along the chain. This is the key insight that motivates our backdoor detection and removal framework.

\noindent
\textbf{Diffusion Models.}
There are three major types of Gaussian-noise-to-image diffusion models: {\em Denoising Diffusion Probabilistic Model} (DDPM)~\cite{Ho.DDPM.NeurIPS.2020}, {\em Noise Conditional Score Network} (NCSN)~\cite{Song.NCSN.NeurIPS.2019}, and {\em Latent Diffusion Model} (LDM)~\cite{Rombach.LDM.CVPR.2022}\footnote{LDM can be considered DDPM in the compressed latent space of a pre-trained autoencoder. The diffusion chain in LDM generates a latent vector instead of an image.}. 
Researchers~\cite{SDE_Diffusion, EDM, Chou.VillanDiff.2023} showed that they can be modeled by a unified Markov Chain denoted in \cref{fig:clean_unified_markov_chain}. From right to left, the forward process $q(x^t|x^{t-1}) = \mathcal{N}(x^t ; \kappa^t x^{t-1}, \upsilon^tI)$ (with $\kappa_t$ denoting transitional content schedulers and $\upsilon^t$  transitional noise schedulers) iteratively adds more noises to a sample $x^0$ until it becomes a Gaussian noise $x^T  \sim \mathcal{N}(0, I)$. 
The training goal of DMs is to learn a network $M_\theta$ to form a reverse process $p_{\theta}(x^{t-1}|x^{t}) = \mathcal{N}(x^{t-1} ;\tilde{\kappa}^t x^{t}+ \hat{\kappa}^t M_{\theta}(x^t, t),\tilde{\upsilon}^tI)$
to iteratively denoise the Gaussian noise $x^T$ to get the sample $x^0$. 
$\hat{\kappa}^t$, $\tilde{\upsilon}^t$, and $\tilde{\upsilon}^t$ are mathematically derived from $\kappa^t$ and $\upsilon^t$.
Different DMs may have different instantiated parameters.

Because the default samplers of DMs are slow~\cite{DDIM,Ho.DDPM.NeurIPS.2020}, 
researchers have proposed different samplers to accelerate it. {\em Denoising Diffusion Implicit Models} (DDIM)~\cite{DDIM}
derives a shorter reverse chain (\textit{e.g.}, 50 steps instead of 1000 steps) from DDPM based on the generalized non-Markovian forward chain. DPM-Solver~\cite{dpm_solver}, 
DPM-Solver++~\cite{dpm_solver_pp}, UniPC~\cite{UniPC}, Heun \cite{EDM}, and DEIS~\cite{DEIS} formulate the diffusion process as an {\em Ordinary Differential Equation} (ODE) and utilize higher-order approximation to generate good samples within fewer steps.

\begin{figure}[t]
	\begin{center}
            \includegraphics[width=0.88\linewidth]{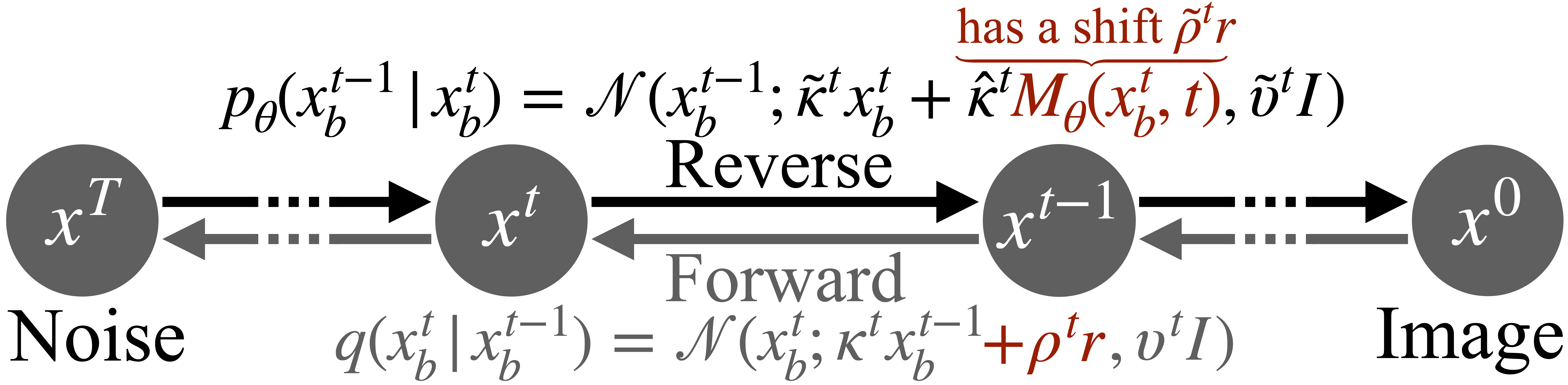}
	\end{center}
	\caption{Backoor injection in DMs via distribution shift.
	}
	\label{fig:backdoored_unified_markov_chain}
\end{figure}

\noindent
\textbf{Backdoor Attacks.}
Different from injecting backdoors to traditional models (e.g., classifiers), which can be achieved solely by data poisoning (i.e., stamping the trigger on the input and forcing the model to produce the intended target output), 
backdoor-attacking DMs is much more complicated as the input space of DMs is merely Gaussian noise and output is an image. 
Specifically, 
attackers first need to carefully and 
\textit{mathematically} define a forward backdoor diffusion process $x^0_b \to x^T_b$ where 
$x^0_b$ is the target image and $x^T_b$ is the input noise with the trigger $r$. As such, a backdoored model can be trained as part of the reverse process. The model generates the target image when $r$ is stamped on an input. To the best of 
our knowledge, there are three existing noise-to-image DM backdoor attacks~\cite{Chou.BadDiff.CVPR.2023, trojdiff, Chou.VillanDiff.2023}. 
Their high-level goal can be 
illustrated in \cref{fig:backdoor_dm_inference}.
Here we use a white square as an example.
When $x^T$ is
a Gaussian noise stamped with a white square 
at the bottom right (the trigger), as shown by the red dotted box on the left, the generated $x^0$ is the target image (i.e., the pink hat on the right).
Inputs with the trigger can be formally defined by a {\em trigger distribution}, i.e., the shifted distribution denoted by the red dotted curve on the left\footnote{The one-dimensional curve is just used to 
conceptually describe the distribution shift. The actual $u_b^T$ is a high dimensional Gaussian with a non zero mean as shown on the left.}.
When 
$x^T \sim \mathcal{N}(0, I)$, $x^0$ is a clean sample.
With the aforementioned unified definition of DMs, we can summarize different attacks as shown in 
\cref{fig:backdoored_unified_markov_chain}.
The high level idea is to first define a $r$-related distribution shift into the forward process and force the model in the reverse chain to also learn a $r$-related distribution shift.
More specifically,
attackers define a backdoor forward process (from right to left at the bottom half) with a distribution shift $\rho^t r$, $\rho^t$ denoting 
the scale of the distribution shift w.r.t $r$
and $r$ the trigger.
During training, their 
backdoor injection objective is to make $M_\theta(x_b^t, t)$'s output at timestep $t$ to shift 
$\tilde{\rho}^tr$ when the input contains the trigger. 
$\tilde{\rho}^t$ denotes the scale of relative distribution shift in the reverse process and is mathematically derived from $\kappa^t$, $\rho^t$ and $\upsilon^t$.
Moreover, the shift at the $x^0$ step is set to produce the target image.
Therefore, during inference, when the input $x^T$ contains the trigger, the backdoored DM will generate the 
target image. Different attacks can be instantiated using different parameters.

\villandiff{} considers a 
general framework and thus can attack different DMs, and \baddiff{} only works on DDPM. 
In \villandiff{} and \baddiff{}, the backdoor reverse process uses the same parameters as the clean one (\textit{i.e.}, the same set of $\tilde{\kappa}^t$, $\hat{\kappa}^t$ and $\tilde{\upsilon}^t$).
\trojdiff{} is slightly different, it focuses on attacking DDPM but needs to manually switch to a separate backdoor reverse process to trigger the backdoor (\textit{i.e.}, a different set of $\tilde{\kappa}^t$, $\hat{\kappa}^t$ and $\tilde{\upsilon}^t$ from the clean one).
It also derives a separate backdoor reverse process to attack DDIM. 
\section{Design}

\begin{figure}[t]
	\begin{center}
            \includegraphics[width=0.8\linewidth]{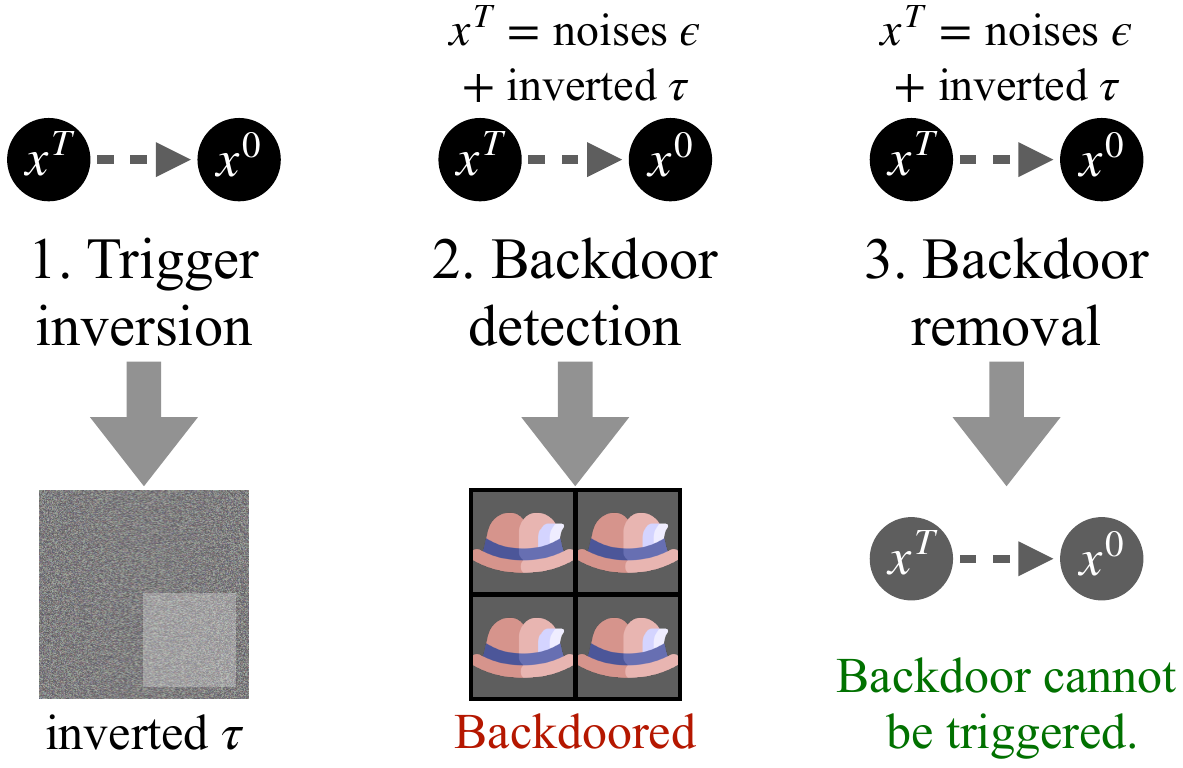}
	\end{center}
	\caption{Workflow of our framework.
	}
	\label{fig:workflow}
\end{figure}

Our framework contains three components as denoted by \cref{fig:workflow}. Given a DM to test, we first run our trigger inversion algorithm to find a potential trigger $\tau$ that has the distribution shift property (\cref{sec:trigger_inversion})\footnote{Here we denote the inverted trigger by $\tau$ to distinguish it from the real trigger $r$ injected by attackers.}. 
Our detection method (\cref{sec:backdoor_detection}) first uses inverted $\tau$ to shift the mean of the input Gaussian distribution to generate a batch of inputs with the trigger. These inputs are fed to the DM to generate a batch of images. Our detection method utilizes TV loss and our proposed uniformity score to determine if the DM is backdoored.
If the DM is backdoored, we run our removal algorithm to eliminate the injected backdoor (\cref{sec:backdoor_removal}).

\begin{figure*}[t]
	\begin{center}
            \includegraphics[width=\linewidth]{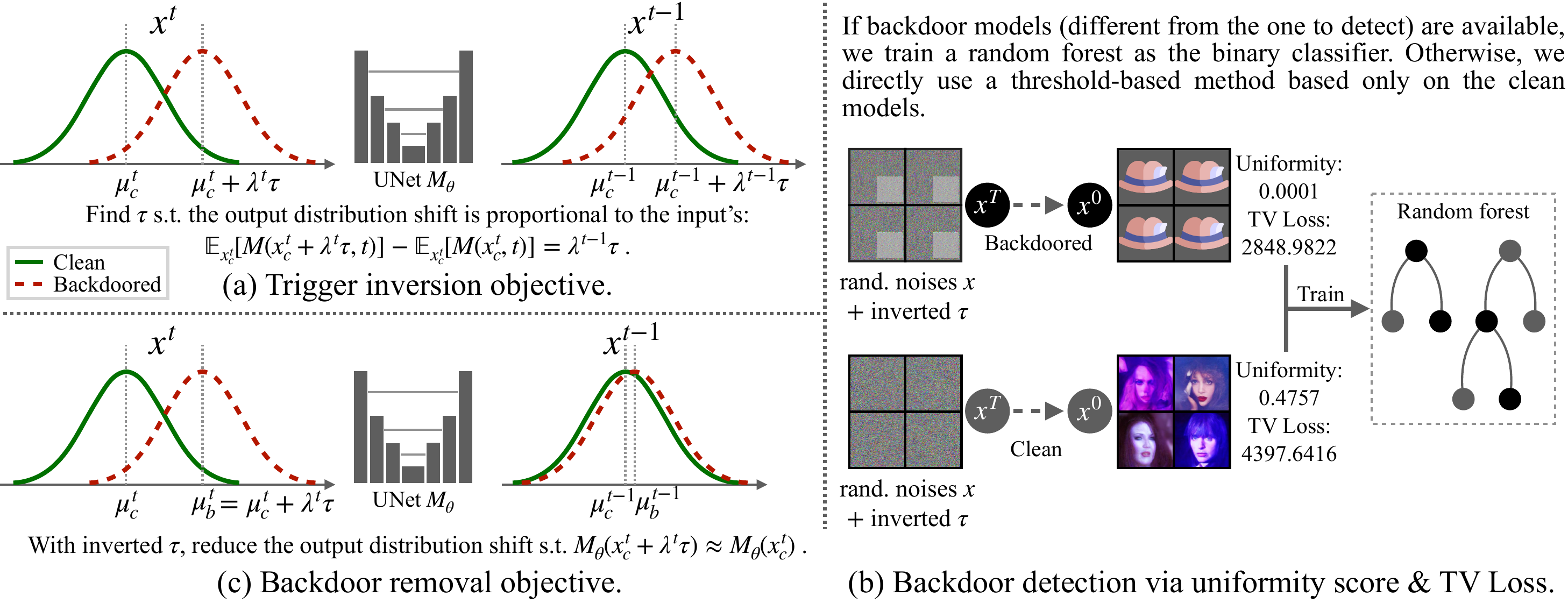}
	\end{center}
	\caption{Overview of our trigger inversion, backdoor detection, and backdoor removal framework.
	}
	\label{fig:overview}
\end{figure*}

\subsection{Trigger Inversion}
\label{sec:trigger_inversion}

Existing trigger inversion techniques focus on discriminative models (such as classifiers and object detectors~\cite{rawat2022devil, baddet} and use classification loss such as the Cross-Entropy loss to invert the trigger. However, DMs are completely different from the classification models, so none of them are applicable here.
As we have seen in \cref{fig:backdoored_unified_markov_chain}, in order to ensure the effectiveness of injected backdoor, attackers need to explicitly preserve the
distribution shift in each timestep along the diffusion chain. In addition, this distribution shift is dependent on the 
trigger input because it is only activated by the trigger.
Therefore, our trigger inversion goal is to find a trigger $\tau$ that can preserve a $\tau$-related shift through the chain.

More specifically, consider at the time step $t$, the noise
$x^t$ is denoised to a less noisy $x^{t-1}$ as denoted in the middle part of \cref{fig:backdoor_dm_inference}.
Denote 
$x^t_c \sim \mathcal{N}(\mu^t_c, \ast)$\footnote{$\ast$ here means our following analysis doesn't consider covariance.} and $x^t_b \sim \mathcal{N}(\mu^t_b, \ast)$ as the noisier clean and backdoor
inputs. Similarly, we use
$x^{t-1}_c \sim \mathcal{N}(\mu^{t-1}_c, \ast)$ and $x^{t-1}_b \sim \mathcal{N}(\mu^{t-1}_b, \ast)$ to denote the less
noisy outputs. As the distribution shift is related to the trigger $\tau$, we model it as a linear dependence and empirically show its effectiveness. 
That is, $\mu^t_b-\mu^t_c = \lambda^t \tau$ and $\mu^{t-1}_b-\mu^{t-1}_c = \lambda^{t-1} \tau$, where
$\lambda^t$ is the coefficient to model the distribution shift relative to $\tau$ at time step $t$.
This leads to our trigger inversion 
objective in \cref{fig:overview} (a). Our goal is to find the trigger $\tau$ that can have the preserved distribution shift, that is\footnote{Note we need no clean samples. The $x^t_c$ in Eq. 1 is transformed from random Gaussian noises $x^T_c \sim \mathcal{N}(0, 1)$.},
\begin{align}
\mathbb{E}_{x^t_c}[M(x^t_c+\lambda^t\tau, t)]-\mathbb{E}_{x^t_c}[M(x^t_c, t)]=\lambda^{t-1}\tau \, , \label{eq:general_loss_tau_t}
\end{align}
Formally,
\begin{align}
\tau &= \arg \min_{\tau} Loss_{\tau} \nonumber \\
Loss_{\tau} &= \mathbb{E}_t [ \| \mathbb{E}_{x^t_c}[M(x^t_c+\lambda^t\tau, t)]-\mathbb{E}_{x^t_c}[M(x^t_c, t)] \nonumber \\
& \phantom{= \mathbb{E}_t [ \|} -\lambda^{t-1}\tau \| ]  \, . \label{eq:general_loss_tau}
\end{align}

A popular way to use the UNet in diffusion models~\cite{Ho.DDPM.NeurIPS.2020, Song.NCSN.NeurIPS.2019} is to predict the added Gaussian noises instead of the noisy images, 
that is $M(x^t_c, t) \sim \mathcal{N}(0, I)$. \cref{eq:general_loss_tau} can be rewritten as
\begin{align}
Loss_{\tau} = \mathbb{E}_t [  \| \mathbb{E}_{x^t_c}[M(x^t_c+\lambda^t\tau, t)] -\lambda^{t-1}\tau \| ] \, . \label{eq:noise_loss_tau}
\end{align}

A straightforward approach to finding $\tau$ is to minimize $Loss_\tau$ computed at each timestep along the chain. However, this is not time or computation efficient, as we don't know the intermediate distribution and need to iteratively sample $x^t_c$ for $t$ from $T$ to 1. Instead, we choose to only consider the timestep $T$ as \cref{eq:general_loss_tau_t} should also hold for $T$. In addition, by definition, we know $x^T_c \sim \mathcal{N}(0, I)$ and $\lambda^T=1$ as $x^T_b \sim \mathcal{N}(\tau, I)$, that is, $\mu^T_b-\mu^T_c = \tau$. Therefore, we can simplify $Loss_\tau$ as 
\begin{align}
Loss_\tau &= \| \mathbb{E}_{x^T_c}[M(x^T_c+\lambda^T\tau, T)] -\lambda^{T-1}\tau \| \nonumber \\
          &= \| \mathbb{E}_{\epsilon \sim \mathcal{N}(0,1)}[M(\epsilon+\tau, T)] -\lambda \tau \| \, , \label{eq:simple_loss_tau}
\end{align}
where we omit the superscript $T-1$ of $\lambda$ for simplicity\footnote{If the UNet is not used to predict the noises, then 
$Loss_\tau = \| \mathbb{E}_{\epsilon \sim \mathcal{N}(0,1)}[M(\epsilon+\tau, T)] - M(\epsilon, T)] -\lambda \tau \|$. In our experiments, we set $\lambda=0.5$ for DDPM.}. 
% \cref{algo:trigger_inversion} 
Algorithm 2
in the appendix shows the pseudocode.

\subsection{Backdoor Detection}
\label{sec:backdoor_detection}

Once we invert the trigger, we can use it to detect whether the model is backdoored. Diffusion models (generative models) are very different from the classifiers. Existing detection methods~\cite{liu2019abs,liu2022complex} on classifiers use the Attack Success Rate (ASR) to measure the effectiveness of the inverted trigger. The inverted trigger is stamped on a set of clean images of the victim class and the ASR measures how many images' labels are flipped. If the ASR is larger than a threshold (such as 90\%), the model is considered backdoored. However, for diffusion models, there are no such label concepts and the target image is unknown. Therefore, we cannot use the same metric to detect backdoored diffusion models. For a similar reason, existing detection methods~\cite{nc} based on the difference in the sizes of the inverted triggers across all labels of a classifier can hardly work either.

As mentioned earlier, the attackers want to generate the target image whenever the input contains the trigger. 
\cref{fig:overview} (b) shows the different behaviors of backdoored and clean diffusion models when the inverted triggers 
$\tau$ are patched to the input noises. For the backdoored model, if the input contains $\tau$, the generated images are the target images. If we know the target image, we can easily compare the similarity (\textit{e.g.}, LPIPS~\cite{Zhang.LPIPS.CVPR.2018}) between the generated
images and the target image. However, we have no such knowledge. Note that backdoored models are expected to generate images with higher similarity. Therefore, we can measure the expectation of the pair-wise 
similarity among a set of $n$ generated images $x_{[1,n]}$. We call it uniformity score:
\begin{align}
    S(x_{[1,n]}) = \mathbb{E}_{i \in [1,n], j \not = i \in [1, n]} [\| x_i - x_j \| ] \, . \label{eq:uniformity_score}
\end{align}

We also compute the average Total Variance Loss, because 1) the target images are not noises, and 2) the inverted trigger usually causes clean models to generate out-of-distribution samples with lower quality since they are not trained with the distribution shift. 
% \cref{algo:feature_extraction} 
Algorithm 3
in the appendix illustrates the feature extraction.

We consider two practical settings to detect a set of models $\mathcal{M}_u$ backdoored by unknown attacks (\textit{e.g.}, \trojdiff{}): 1) we have access to a set of backdoored models $\mathcal{M}_b$ attacked by a different method (\textit{e.g.}, \baddiff{}) and a set of clean models $\mathcal{M}_c$, or 2) we only can access a set of clean models $\mathcal{M}_c$.

In the first setting, these two features extracted for $\mathcal{M}_b$ and $\mathcal{M}_c$ with the corresponding labels are used to train a random forest as the backdoor detector to detect $\mathcal{M}_b$\footnote{We choose the random forest because it doesn't require input normalization. TV loss value range is much larger than the uniformity score as shown in \cref{fig:uniformity_tvloss}.}.
% \cref{algo:backdoor_detection_random_forest} 
Algorithm 4
shows backdoor detection in this setting.

In the second setting, we extract one feature for $\mathcal{M}_c$ and compute a threshold for each feature based on a selected false positive rate (FPR) such as 5\%, meaning that our detector classifies 5\% of clean models as trojaned using the threshold.  For a model in $\mathcal{M}_u$, if its feature value is smaller than the threshold, it's considered backdoored.
Note that due to the nature of these thresholds, the high FPR we tolerate, the fewer truly trojaned models we may miss.
The procedure is described in 
Algorithm 5.
% \cref{algo:backdoor_detection_threshold}.

\subsection{Backdoor Removal}
\label{sec:backdoor_removal}

Trigger inverted and backdoor detected, we can prevent the jeopardy by not deploying the backdoored model. However, this also forces us to discard the learned benign utility. Therefore, we devise an approach to removing the injected backdoor while largely maintaining the benign utility.

Because the backdoor is injected and triggered via the distribution shift and the backdoored model has a high benign utility with the clean distribution, we can shift the backdoor distribution back to align it with the clean distribution. Its objective is demonstrated in \cref{fig:overview} (c). Formally, given the inverted trigger $\tau$ and the backdoored model $M_\theta$, our goal is to minimize the following loss:    
\begin{align}
    Loss_{rb} = \mathbb{E}_t [\mathbb{E}_{x^t_c}[\| M_\theta(x^t_c+ \lambda^t \tau)-M_\theta(x^t_c) \|]] \, .
\end{align}

Similar to trigger inversion loss, we can apply $Loss_{rb}$ only at the timestep $T$ and simplify it as:
\begin{align}
    Loss_{rb} =  \mathbb{E}_{\epsilon \sim \mathcal{N}(0, 1)}[\| M_\theta(\epsilon + \tau)-M_\theta(\epsilon) \|] \, .
\end{align}

However, this loss alone is not sufficient, because $M_\theta$ may learn to shift the benign distribution towards the backdoor one instead of the opposite.
Therefore, we use
the clean distribution of $M_\theta$ on clean inputs as a reference. To avoid interference, we clone $M_\theta$ and freeze the clone's weights. The 
frozen model is denoted as $M_f$ and $Loss_{rb}$ is changed to:
\begin{align}
    Loss_{rb} =  \mathbb{E}_{\epsilon \sim \mathcal{N}(0, 1)}[\| M_\theta(\epsilon + \tau)-M_f(\epsilon) \|] \, .
\end{align}

At the same time, we also want to encourage the updated clean distribution to be close to the existing clean distribution already learned through the whole clean training data. It can be expressed as:
\begin{align}
    Loss_{mc} = \mathbb{E}_{\epsilon \sim \mathcal{N}(0, 1)}[\| M_\theta(\epsilon)-M_f(\epsilon) \|] \, .
\end{align}

With $Loss_{rb}+Loss_{mc}$, we can get $M_{\theta'}$ to invalidate injected backdoor and the ground truth trigger very fast in 20 updates as shown by 
Figure 9
% \cref{fig:only_backdoor_removal_loss} 
in the appendix~\cite{an2023elijah_arxiv}. That is, when we feed the input noise patched with the ground truth trigger, $M_{\theta'}$ won't generate the target image. However, 
% \cref{algo:trigger_inversion} 
Algorithm 2
can invert another trigger $\tau'$ that can make $M_{\theta'}$ output the target image. 
A plausible solution is to train with more iterations.
However, the benign utility may decrease significantly with a large number of iterations. So we add the original clean training loss $Loss_{dm}$\footnote{This is the vanilla training loss for DMs on clean data. Please refer to the appendix or the original papers for more details.} of diffusion models into our backdoor removal procedure. There are two ways to use $Loss_{dm}$. The first way follows existing backdoor removal literature~\cite{liu2018fine,borgnia2020strong,zeng2020deepsweep,moth}, where we can access 10\% clean training data. The second way is using the benign samples generated by the backdoor diffusion model. Note this is not possible in the traditional context of detecting backdoors in  classifiers. Hence, our complete loss for backdoor removal is
\begin{align}
    Loss_\theta = Loss_{rb} + Loss_{mc} + Loss_{dm} \, . \label{eq:backdoor_removal_loss_theta}
\end{align}

Our backdoor removal method is described by 
Algorithm 6
% \cref{algo:backdoor_removal} 
in the appendix~\cite{an2023elijah_arxiv}.
\section{Evaluation}
\label{sec:evaluation}

We implement our framework \ours{} including trigger inversion, backdoor detection, and backdoor removal algorithms in PyTorch~\cite{Facebook.PyTorch.NeurIPS.2019}\footnote{Our code: \url{https://github.com/njuaplusplus/Elijah}}. We evaluate our methods on hundreds of models including the three major architectures and different samplers against all published backdoor attack methods for diffusion models in the literature. Our main experiments run on a server equipped with Intel Xeon Silver 4214 2.40GHz 12-core CPUs with 188 GB RAM and NVIDIA Quadro RTX A6000 GPUs.

\subsection{Experimental Setup}

\noindent
\textbf{Datasets, Models, and Attacks.}
We mainly use the CIFAR-10~\cite{cifar10} and downscaled CelebA-HQ~\cite{karras2018progressive} datasets as they are the two datasets considered in the evaluated backdoor attack methods.
The CIFAR-10 dataset contains 60K $32\times32$ images of 10 different classes, while the CelebA-HQ dataset includes 30K faces resized to $256\times256$. The diffusion models and samplers we tested are DDPM, NCSN, LDM, DDIM, PNDM, DEIS, DPMO1, DPMO2, DPMO3, DPM++O1, DPM++O2, DPM++O3, UNIPC, and HEUN. Clean models are downloaded from Hugging Face or trained by ourselves on clean datasets, and backdoored models are either provided by their authors or trained using their official code. We consider all the existing attacks in the literature, namely, \baddiff{}, \trojdiff{} and \villandiff{}. In total, we use 151 clean and 296 backdoored models. Details of our configuration can be found in Appendix B.1~\cite{an2023elijah_arxiv}.

\noindent
\textbf{Baselines}
To the best of our knowledge, there is no backdoor detection or removal techniques proposed for DMs. So we don't have direct baselines. Also, as we mentioned earlier, most existing defenses are specific for discriminative models and thus are not applicable to generative models.

\noindent
\textbf{Evaluation Metrics.} We follow existing literature~\cite{liu2019abs,Chou.BadDiff.CVPR.2023,trojdiff,Chou.VillanDiff.2023} and use the following metrics:
1) \textbf{Detection Accuracy (ACC)} assesses the ability of our backdoor detection method. It's the higher the better. 
Note that the training and testing datasets contain \textit{non-overlapping attacks} in the setting where we assume backdoored models are available for training.
2) \textbf{$\Delta$FID} measures the relative Fréchet Inception Distance (FID) \cite{FID} changes between the backdoored model and the fixed model.
It shows our effects on benign utility. It's the smaller the better.
3) \textbf{$\Delta$ASR} denotes the relative change in Attack Success Rate (ASR) and shows how well our method can remove the backdoor. ASR calculates the percentage of images generated with the trigger input that are similar enough to the target image (\textit{i.e.}, the MSE w.r.t the target image is smaller than a pre-defined threshold~\cite{Chou.VillanDiff.2023}). A smaller $\Delta$ASR means a better backdoor removal.
4) \textbf{$\Delta$SSIM} also evaluates the effectiveness of the backdoor removal, similar to $\Delta$ASR. It computes the relative change in Structural Similarity Index Measure (SSIM) before and after the backdoor removal. A Higher SSIM means the generated image is more similar to the target image. Therefore, a smaller $\Delta$SSIM implies a better backdoor removal.

\begin{figure}[t]
	\begin{center}
            \includegraphics[width=\linewidth]{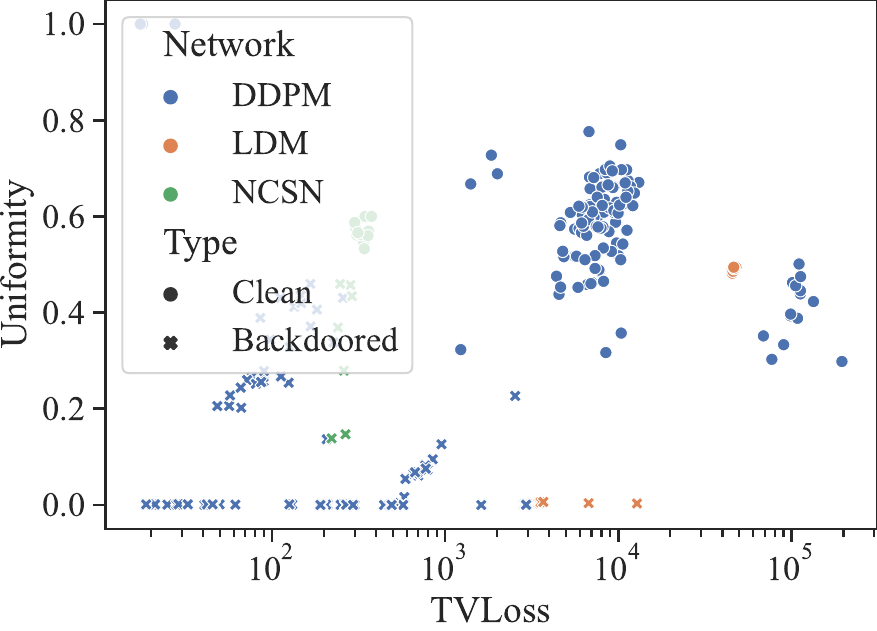}
	\end{center}
	\caption{Uniformity scores and TVLoss for 151 clean and 296 backdoored models.
	}
	\label{fig:uniformity_tvloss}
\end{figure}

\begin{table}[t]
    % \footnotesize
    \begin{center}
    % \scriptsize
    \tabcolsep=2pt
    \begin{tabular}{cccccc}
        \toprule
        Attack & Model & ACC$\uparrow$ & $\Delta$ASR$\downarrow$ & $\Delta$SSIM$\downarrow$ & $\Delta$FID$\downarrow$   \\
        \midrule
        \multicolumn{2}{c}{Average} & 1.00	& -0.99	& -0.97	& 0.03\\
        \midrule
        \baddiff{} & DDPM-C       & 1.00   & -1.00 & -0.99 & -0.00 \\
        \baddiff{} & DDPM-A       & 1.00   & -1.00 & -1.00 & 0.10  \\
        \trojdiff{} & DDPM-C      & 0.98 & -1.00 & -0.96 & 0.04  \\
        \trojdiff{} & DDIM-C      & 0.98 & -1.00 & -0.96 & 0.03  \\
        \villandiff{} & NCSN-C    & 1.00   & -0.96 & -0.90 & 0.17  \\
        \villandiff{} & LDM-A     & 1.00   & -1.00 & -0.99 & -0.31 \\
        \villandiff{} & ODE-C     & 1.00   & -1.00 & -1.00 & 0.17  \\

        \bottomrule
    \end{tabular}
    \caption{Overall results of backdoor detection and removal. Model DDPM-C (resp. DDPM-A) means DDPM models trained on CIFAR-10 (resp. CelebA-HQ) dataset. Here ODE-C shows the average results for ODE samplers attacked by \villandiff{}. 
    % Results of samplers are in 
    % \cref{tab:ode_results}.
    }
    \label{tab:overall_results}
    \end{center}
\end{table}

\subsection{Backdoor Detection Performance}

Our backdoor detection uses the uniformity score and TV loss as the features. \cref{fig:uniformity_tvloss} shows the 
distribution of clean and backdoored models in the extracted feature space. Different colors denote different networks. The circles denote clean models while the crosses are backdoored ones. The two extracted features are very informative as we can see clean and backdoored models are quite separable.
The third column of \cref{tab:overall_results} reflects the detection accuracy when we can access models backdoored by attacks different from the one to detect.
Our average detection accuracy is close to 100\%, and in more than half of the cases, we have an accuracy of 100\%. Our detection performance with only access to clean models is comparable and shown in Table 2
% \cref{tab:detection_threshold} 
in Appendix B.2~\cite{an2023elijah_arxiv}. 
% \cref{fig:inverted_trigger} 
Figure 14 in Appendix B.9~\cite{an2023elijah_arxiv} visualizes some inverted triggers.

\subsection{Backdoor Removal Performance}

The last three columns in \cref{tab:overall_results} show the overall results. As shown by the $\Delta$ASR, we can remove the injected backdoor completely for all models except for NCSN (almost completely). $\Delta$SSIM reports similar results. With the trigger input, the images generated by the backdoored models have high SSIM with the target images, while after the backdoor removal, they cannot generate the target images. The model utility isn't significantly sacrificed as the average $\Delta$FID is 0.03. For some FIDs with nontrivial increases, the noise space and models are larger (DDPM-A), or the models themselves are more sensitive to training on small datasets (NCSN-C and ODE-C). Detailed ODE results are in Appendix B.8~\cite{an2023elijah_arxiv}.

\subsection{Effect of Backdoor Removal Loss}
\label{sec:effect_of_backdoor_removal_loss}

\begin{figure}[t]
	\begin{center}
            \includegraphics[width=\linewidth]{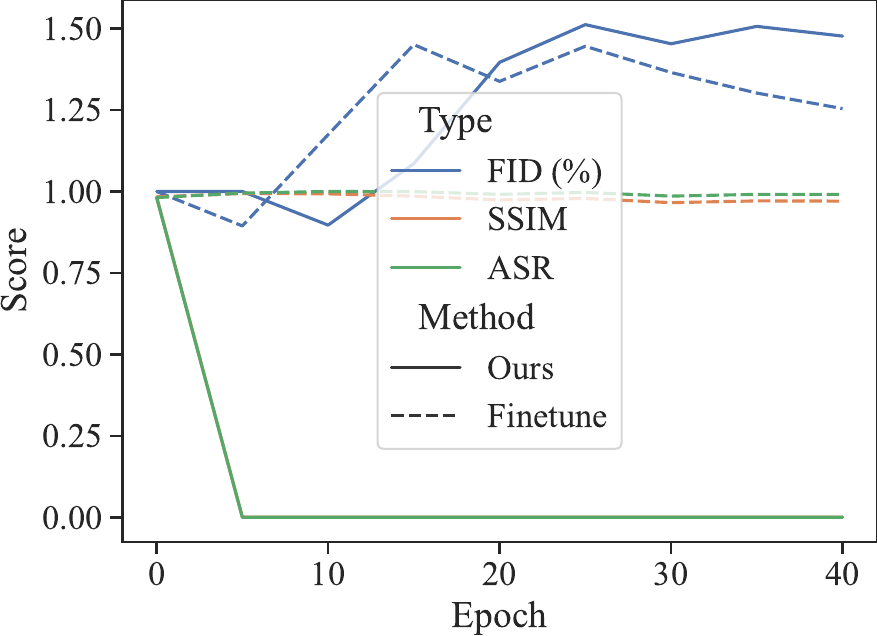}
	\end{center}
	\caption{Fine-tuning with only real data cannot remove the injected backdoor. The model is backdoored by \baddiff{} with the stop sign trigger and the hat target.
	}
	\label{fig:ours_finetune_stopsign_hat}
\end{figure}

\cref{fig:ours_finetune_stopsign_hat} shows fine-tuning the backdoored model only with 10\% clean data cannot remove the backdoor. The green dashed line displays the ASR which is always close to 1 for the fine-tuning method, while ours (denoted by the solid green line) quickly reduces ASR to 0 within 5 epochs. More results can be found in Appendix B.6~\cite{an2023elijah_arxiv}. Our appendix also studies the effects of other factors such as the trigger size (Appendix B.3), the poisoning rate (Appendix B.4), and clean data (Appendix B.5).

\subsection{Backdoor Removal with Real/Synthetic Data}

\begin{figure}[t]
	\begin{center}
            \includegraphics[width=\linewidth]{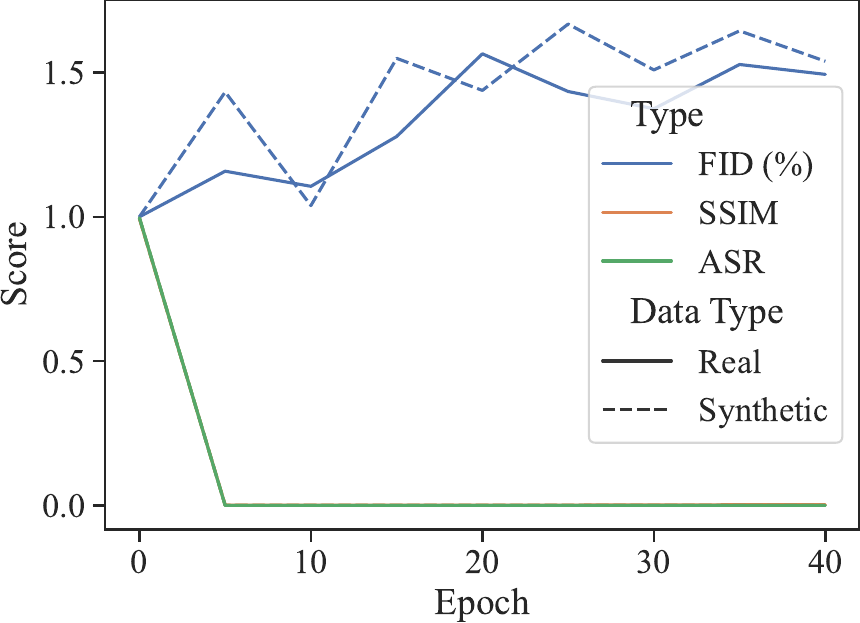}
	\end{center}
	\caption{Backdoor removal with real data or synthetic data on a model backdoored by \baddiff{} with the stop sign trigger and the box target. ASR and SSIM lines overlap.
	}
	\label{fig:gt_synthetic_comparison}
\end{figure}

One advantage of backdoor removal in DMs over other models (\textit{e.g.}, classifiers) is we can use the DMs to generate synthetic data instead of requiring ground truth clean training data. This is based on the fact that backdoored models also maintain high clean utility. \cref{fig:gt_synthetic_comparison} shows how \ours{} performs with 10\% real data or the same amount of synthetic data. The overlapped SSIM and ASR lines mean the same effectiveness of backdoor removal. The FID changes in a similar trend. This means we can have a real-data-free backdoor removal approach. Since our backdoor detection is also sample-free, our whole framework can work even without access to real data. 
% More results can be found in Appendix B.7~\cite{an2023elijah_arxiv}.

\subsection{Adaptive Attacks}

We evaluate our framework against the strongest attackers who know our framework. Therefore, in order to make the injected backdoor undetectable and irremovable by our framework, they add our backdoor removal loss into their backdoor attack loss. However, the backdoor injection is not successful. This is expected because the backdoor attack loss contradicts our backdoor removal loss. More details can be found in the appendix~\cite{an2023elijah_arxiv}.

\section{Related Work}
\label{sec:related_work}

\noindent
\textbf{Diffusion Models and Backdoors.}
Diffusion Models have attracted a lot of researchers, to propose different models~\cite{Ho.DDPM.NeurIPS.2020,Song.NCSN.NeurIPS.2019,NCSNv2,EDM,SDE_Diffusion,Rombach.LDM.CVPR.2022} and different applications~\cite{GeoDiff, Diff-TTS, Grad-TTS, Guided-TTS, DiffWave, VIDM, VideoDiff,DreamBooth.CVPR.2023}. A lot of methods are proposed to deal with the slow sampling process~\cite{DDIM,dpm_solver,dpm_solver_pp,UniPC,EDM,DEIS}
Though they achieve a huge success, they are vulnerable to backdoor attacks~\cite{Chou.BadDiff.CVPR.2023,trojdiff,Chou.VillanDiff.2023}. To mitigate this issue, we propose the first backdoor detection and removal framework for diffusion models.

% \smallskip
\noindent
\textbf{Backdoor Attacks and Defenses.} When backdooring discriminative models, some poison labels~\cite{blend,trojnn} while others use clean label~\cite{clean-label, hidden}. 
These attacks can manifest across various modalities~\cite{lock-nlp, backdoorl}. 
Backdoor defense encompasses backdoor scanning on model and dataset~\cite{liu2019abs,featureRE} 
and certified robustness~\cite{patchcleanser,jia2022certified}.
Backdoor removal aims to detect poisoned data through techniques like data sanitization~\cite{tang2019demon,gao2019strip} and to eliminate injected backdoors from contaminated models~\cite{nc, pixel, moth, flip, medic}. 
These collective efforts highlight the critical importance of defending against backdoor threats in the evolving landscape of machine learning security. However, existing backdoor defense mechanisms and removal techniques have not been tailored to the context of diffusion models. 
Appendix E has an example~\cite{an2023elijah_arxiv}.

\section{Conclusion}
\label{sec:conclusion}

We observe existing backdoor attacks in DMs utlize distribution shift and propose the first backdoor detecion and removal framework \ours{}. Extensive experiments show \ours{} can have a close-to-100\% detection accuracy, eliminate backdoor effects completely in most cases without significantly sacrificing the model utility.

\section*{Acknowledgments}
We thank the anonymous reviewers for their constructive comments. We are grateful to the Center for AI
Safety for providing computational resources. This research
was supported, in part by IARPA TrojAI W911NF-19-S0012, NSF 1901242 and 1910300, ONR N000141712045,
N000141410468 and N000141712947. Any opinions, findings, and conclusions in this paper are those of the authors
only and do not necessarily reflect the views of our sponsors. Shengwei would like to express his deepest appreciation to his wife Jinyuan and baby Elijah.

\bibliography{aaai24}

\appendix
% \section*{Appendix}
\begin{center}
\LARGE \bf Appendix
\end{center}

\begin{algorithm}[htbp]
  % \small
  \caption{Overall algorithm of \ours{}.}
  \label{algo:overall_algo}
  \begin{algorithmic}[1]
  \Function{\ours{}}{diffusion model $Dm$}
    \State $\tau =$ \Call{InvertTrigger}{$Dm.M_{\theta}$}
    \If{\Call{DetectBackdoor}{$Dm$, $\tau$}}
      \State $Dm.M_{\theta} = $ \Call{RemoveBackdoor}{$Dm.M_{\theta}$, $\tau$}
    \EndIf
    \State \Return $Dm$
  \EndFunction
  \end{algorithmic}
\end{algorithm}

\begin{algorithm}[t]
  % \small
  \caption{Trigger inversion. $T$ is the leftmost step. }
  \label{algo:trigger_inversion}
  \begin{algorithmic}[1]
  \Function{InvertTrigger}{model $M$, epoch $e$, lr $\eta$}
    \State Init $\tau$ from $\mathcal{U}[0, 1)$
    \For {$i \in \{1, \ldots, e\}$}
      \State Sample $\epsilon$ from $\mathcal{N}(0, I)$
      \State $\tau = \tau - \eta \nabla_{\tau} Loss_{\tau}(\epsilon)$
      \Comment{\cref{eq:simple_loss_tau}}
    \EndFor
    \State \Return $\tau$
  \EndFunction
  \end{algorithmic}
\end{algorithm}
% \vspace{20pt}
% \vspace{-10pt}
% \setlength{\textfloatsep}{10pt}
\begin{algorithm}[t]
  % \small
  \caption{Extract features for backdoor detection.}
  \label{algo:feature_extraction}
  \begin{algorithmic}[1]
  \Function{ExtractFeature}{diffusion model $Dm$, img\_num $n$, epoch $e$, lr $\eta$}
    \State $\tau =$ \Call{InvertTrigger}{$Dm.M$, $e$, $\eta$}
    \State Sample $x^T_{b, [1,n]}$ from $\mathcal{N}(\tau, I)$
    \State Generate $n$ images $x_{[1, n]} = Dm(x^T_{b, [1,n]}) $
    \State Compute uniformity $s = S(x_{[1, n]})$ \Comment{\cref{eq:uniformity_score}}
    \State Compute TV Loss $ l = L_{TV}(x_{[1, n]})$
    \State \Return $(s, l)$
  \EndFunction
  \end{algorithmic}
\end{algorithm}

\begin{algorithm}[t]
  % \small
  \caption{Backdoor detection via random forest.}
  \label{algo:backdoor_detection_random_forest}
  \begin{algorithmic}[1]
  \Function{BuildDetector}{diffusion models $\{Dm_i\}$, labels $\{l_i \in \{$ `c', `b' $\} \}$, img\_num $n$, epoch $e$, lr $\eta$}
    \State $\mathcal{D}_{train} = \{ \}$
    \For {$ Dm_i, l_i \in \{Dm_i\}, \{l_i\}$}
      \State $f_i = \text{\Call{ExtractFeature}{$Dm_i$, $n$, $e$, $lr$}}$
      \State $\mathcal{D}_{train} = \mathcal{D}_{train} \cup \{( f_i, l_i)\}$
    \EndFor
    \State Train a random forest classifier $cls$ on $\mathcal{D}_{train}$
    \State \Return $cls$
  \EndFunction

  \Function{DetectBackdoor}{classifier $cls$, diffusion model $Dm$, img\_num $n$, epoch $e$, lr $\eta$}
    \State  $f = \text{\Call{ExtractFeature}{$Dm$, $n$, $e$, $lr$}}$
    \State \Return $cls(f)$
  \EndFunction
  \end{algorithmic}
\end{algorithm}
% \vspace{20pt}
% \vspace{-10pt}

\begin{algorithm}[t]
  % \small
  \caption{Backdoor detection via threshold.}
  \label{algo:backdoor_detection_threshold}
  \begin{algorithmic}[1]
  \Function{ComputeThreshold}{clean diffusion models $\{Dm_i\}$, img\_num $n$, epoch $e$, lr $\eta$, FPR $\gamma$}
    \State $\mathcal{U}_{train} = [\ ]$, $\mathcal{T}_{train} = [\ ]$
    \For {$ Dm_i \in \{Dm_i\}$}
      \State $s_i, l_i = \text{\Call{ExtractFeature}{$Dm_i$, $n$, $e$, $lr$}}$
      \State $\mathcal{U}_{train}$.append$( s_i)$
      \State $\mathcal{T}_{train}$.append$(l_i)$
    \EndFor
    \State Sort $\mathcal{U}_{train}$ and $\mathcal{T}_{train}$ in ascending order
    \State $\psi_U = \mathcal{U}_{train}[\lfloor |\mathcal{U}_{train}| * \gamma \rfloor]$
    \State $\psi_T = \mathcal{T}_{train}[\lfloor |\mathcal{T}_{train}| * \gamma \rfloor]$
    \State \Return $\psi_U$, $\psi_T$
  \EndFunction

  \Function{DetectBackdoorU}{diffusion model $Dm$, threshold $\psi$, img\_num $n$, epoch $e$, lr $\eta$}
    \State  $s = \text{\Call{ExtractFeature}{$Dm$, $n$, $e$, $lr$}}[0]$
    \State \Return $s < \psi$
  \EndFunction

  \Function{DetectBackdoorT}{diffusion model $Dm$, threshold $\psi$, img\_num $n$, epoch $e$, lr $\eta$}
    \State  $l = \text{\Call{ExtractFeature}{$Dm$, $n$, $e$, $lr$}}[1]$
    \State \Return $l < \psi$
  \EndFunction
  \end{algorithmic}
\end{algorithm}
\begin{algorithm}[t]
  % \small
  \caption{Backdoor removal. $T$ is the leftmost step. }
  \label{algo:backdoor_removal}
  \begin{algorithmic}[1]
  \Function{RemoveBackdoor}{model $M_{\theta}$, epoch $e$, lr $\eta$, trigger $\tau$, clean data $\mathcal{D}$}
    \State $M_f =$ \Call{Freeze}{$M_{\theta}$}
    \For {$i \in \{1, \ldots, e\}$}
      \State Sample $x^0$ from $\mathcal{D}$
      \State Sample $x^T$ from $\mathcal{N}(0, I)$
      \State $\epsilon_{c} = M_{\theta}(x^T, T)$
      \State $\epsilon_{b} = M_{\theta}(x^T+\tau, T)$
      \State $\epsilon_{f} = M_f (x^T, T)$
      % \State $\ell = Loss_{\epsilon}(\epsilon_{clean}, \epsilon_{freeze}) + Loss_{\epsilon}(\epsilon_{backdoor}, \epsilon_{freeze})$
      \State $\theta = \theta - \eta \nabla_{\theta} Loss_{\theta}(\epsilon_{c}, \epsilon_{b}, \epsilon_{f}, x^0)$ \Comment{Eq. (\ref{eq:backdoor_removal_loss_theta})}
    \EndFor
    \State \Return $M_{\theta}$
  \EndFunction
  \end{algorithmic}
\end{algorithm}
% \vspace{20pt}
% \vspace{-10pt}

\section{Algorithms}

Note that for simplicity, we assume the tensors (\textit{e.g.}, $\epsilon$, $x^T$, etc.) can be a batch of samples or inputs. We only add the indices (\textit{e.g.}, $x^T_{[1, n]}$) when we emphasize the number of samples (\textit{e.g.}, $n$). We call the whole chain a diffusion model denoted by $Dm$ and the learned network model $M$. $Dm(x^T)$ will generate the final image $x^0$ iteratively calling $M$ for $T$ steps.

\cref{algo:overall_algo} shows the overall algorithm of our framework. For simplicity, we omit other parameters here. Because our original \textsc{DetectBackdoor} (\cref{algo:backdoor_detection_random_forest,algo:backdoor_detection_threshold}) calls \textsc{InvertTrigger} (\cref{algo:trigger_inversion}), it doesn't need $\tau$ as a parameter. Here we explicitly run \textsc{InvertTrigger} and pass  $\tau$  to \textsc{DetectBackdoor} to illustrate the workflow. Details of each sub-algorithm are explained as follows.

\subsection{Trigger Inversion}

\cref{algo:trigger_inversion} shows the pseudocode of our trigger inversion. For a given diffusion model to test, we feed its learned network $M$ to \cref{algo:trigger_inversion}, and set the epochs and learning rate. Line 2 initializes $\tau$ using a uniform distribution. Line 3-6 iteratively update $\tau$ using the gradient descent on the trigger inversion loss defined in \cref{eq:simple_loss_tau}. Line 7 returns the inverted $\tau$.

\subsection{Backdoor Detection}

\cref{algo:feature_extraction} describes how we extract the uniformity score and TV loss for a diffusion model $Dm$. Line 2 calls \cref{algo:trigger_inversion} to invert a trigger $\tau$ for the learned network $Dm.M$. Line 3 samples a set of $n$ trigger inputs $x^T_{b, [1,n]}$ from the $\tau$-shifted distribution $\mathcal{N}(\tau, I)$. Line 4 generates a batch of images $x_{[1,n]}$ by feeding $x^T_{b, [1,n]}$ to $Dm$. Lines 5 and 6 compute uniformity score and TV loss. Line 7 returns the extracted features.

\cref{algo:backdoor_detection_random_forest} shows how we conduct backdoor detection when we have both clean models and backdoored models (attacked by a different method from the one to detect). \textsc{BuildDetector} takes in a set of clean and backdoored models $\{Dm_i\}$ and the corresponding labels $\{l_i \in \{$ `c', `b' $\} \}$ where `c' stands for clean and `b' for backdoored. Lines 2-6 extract features for all models and build the training dataset $\mathcal{D}_{train}$ for the random forest. Lines 7-8 train and return a learned random forest classifier $cls$. Given a model to test, \textsc{DetectBackdoor} extracts its features (Line 11) and uses $cls$ to predict its label (Line 12).

When we only have access to the clean models, we use threshold-based detection in \cref{algo:backdoor_detection_threshold}. \textsc{ComputeThreshold} computes the thresholds for the uniformity score and TV loss. More specifically, Line 2 initializes the two empty feature lists. Lines 3-6 compute the uniformity score and TV loss for each clean DM and add them to the lists. Line 8 sorts the two lists in ascending order. Line 9-10 computes the thresholds for uniformity score and TV loss according to the defined False Positive Rate (FPR). Line 11 returns the thresholds. The threshold-based detection algorithms \textsc{DetectBackdoorU} and \textsc{DetectBackdoorT} are straightforward. Given a model to check, they compute the feature uniformity score (Line 14) or TV loss (Line 18). If the feature is smaller than the threshold, the model is considered backdoored (Line 15 or 16).

\subsection{Backdoor Removal}

\cref{algo:backdoor_removal} shows the procedure of removing the backdoor. Given a 
backdoored model $M_\theta$, the inverted trigger $\tau$ and a set of clean (real or 
synthetic) data $\mathcal{D}$, Line 2 first gets a frozen copy of the backdoored model.
Lines 3-9 apply the backdoor removal loop for $e$ epochs. Line 4 gets the clean samples from $D$ as the training samples. Line 5 samples the initial clean noise $x^T$. Lines 6 and 7 compute the backdoored model's outputs of clean inputs and backdoored ones. Line 8 computes the frozen model's outputs of clean inputs as the reference. Line 9 updates $\theta$ using gradient descent on our backdoor removal loss. Line 11 returns the fixed model.

\section{More Experimental Details}

\subsection{Configuration}

\noindent
\textbf{Runs of Algorithms.}
For each model in the 151 clean and 296 backdoored models, we run \cref{algo:trigger_inversion} and \cref{algo:backdoor_removal} once, except for \cref{algo:backdoor_detection_random_forest} and \cref{algo:backdoor_detection_threshold} because clean models and some backdoored models are involved in multiple detection experiments. Given the number of models we evaluated, we believe the results should be reliable.

\noindent
\textbf{Parameters and Settings.}
For trigger inversion, we use Adam optimizer~\cite{Kingma.Adam.ICLR.2015} and 0.1 as the learning rate. We use 100 epochs for DDPM/NCSN, 10 epochs for LDM and ODE models. We set the batch size to 100 for DMs with $3\times32\times32$ space, 50 for $3\times128\times128$ space, and 20 for $3\times256\times256$ space because of GPU memory limitation. Ideally, a larger batch size will give us a better approximation of the expectation in \cref{eq:simple_loss_tau}. We set $\lambda=0.5$ because we tested on a subset of models for $\lambda \in [0, 1]$ with steps 0.1 and found $\lambda=0.5$ gave the best detection results.

For feature extraction, we only use 16 images generated by input with the inverted trigger since we find it's sufficient.

For the random-forest-based backdoor detection, we randomly split the clean model into 80\% training and 20\% testing. We add all the backdoored models by one attack to the test dataset. We add all the backdoored models attacked by a different method from the one to test into the training data. 

For the threshold-based backdoor detection, we split the clean model into 80\% training and 20\% testing. We add all the backdoored models by one attack to the test dataset. We derive the thresholds based on the clean training dataset.

To compute $\Delta$ASR, $\Delta$SSIM, and $\Delta$FID, we use 2048 generated images. Generating a lot of images for hundreds of models is very consuming. For example, it take more than 2 hours to generate 2048 $32\times32$ samples using NCSN trained on the Cifar10 dataset with batch size 2048 on a NVIDIA Quadro RTX A6000 GPU. Since we are comparing the changes, the trends can be implied by using the same reasonable amount of samples to compare the metric on backdoored model and the corresponding fixed one.

\subsection{Threshold-based Detection}

\begin{table}[t]
    % \footnotesize
    \begin{center}
    \caption{Detection accuracy with different settings. ACC means the detection rate with the trained random forest. U@05 means using the threshold extracted on the cleaning training set with a 5\% false positive rate. Model DDPM-C (resp. DDPM-A) means DDPM models trained on CIFAR-10 (resp. CelebA-HQ) dataset.}
    \label{tab:detection_threshold}
    % \scriptsize
    \tabcolsep=2pt
    \begin{tabular}{ccccc}
        \toprule
        Attack & Model & ACC(\%)$\uparrow$ & U@5(\%)$\uparrow$ & T@5(\%)$\uparrow$   \\
        \midrule
        \baddiff{} & DDPM-C       & 100   &  92.04 &  98.23 \\
        \baddiff{} & DDPM-A       & 100   &  100   &  100   \\
        \trojdiff{} & DDPM-C      & 98.36 &  100   &  96.36 \\
        \trojdiff{} & DDIM-C      & 98.36 &  100   &  96.36 \\
        \villandiff{} & NCSN-C    & 100   &  100   &  100   \\
        \villandiff{} & LDM-A     & 100   &  100   &  100   \\
        \villandiff{} & ODE-C     & 100   &  100   &  98.50 \\
        \bottomrule
    \end{tabular}
    \end{center}
\end{table} 
\begin{table}[t]
    % \footnotesize
    \begin{center}
    \caption{Performance against different trigger sizes. Numbers show the relative scores compared with backdoored models. The trigger is a white square and the target image is mickey.}
    \label{tab:different_trigger_sizes}
    % \scriptsize
    \tabcolsep=2.5pt
    \begin{tabular}{ccccc}
        \toprule
        Trigger size & Detected & $\Delta$ASR$\downarrow$ & $\Delta$SSIM$\downarrow$ & $\Delta$FID$\downarrow$ \\
        \midrule
        3$\times$3 & \ding{52} & -1.00 & -0.97 & 0.02 \\
        4$\times$4 & \ding{52} & -1.00 & -0.97 & 0.04 \\
        5$\times$5 & \ding{52} & -1.00 & -0.92 & 0.08 \\
        6$\times$6 & \ding{52} & -1.00 & -0.94 & 0.11 \\
        7$\times$7 & \ding{52} & -1.00 & -0.94 & 0.05 \\
        8$\times$8 & \ding{52} & -1.00 & -0.93 & 0.02 \\
        9$\times$9 & \ding{52} & -1.00 & -0.92 & 0.05 \\
        % \midrule
        \bottomrule
    \end{tabular}
    \end{center}
\end{table}

\cref{tab:detection_threshold} compares our detection performance between the random-forest-based method and the threshold-based one with the false positive rate set to 5\%. The third column shows the detection accuracy with the random forest. The fourth/fifth column shows the results with a uniformity/TV loss threshold. They perform comparably well while the random forest have a overall higher accuracy.

\subsection{Effect of the Trigger Size}

We use \trojdiff{} to backdoor DMs with various trigger sizes and test \ours{} on them. Results are shown in \cref{tab:different_trigger_sizes}. \ours{} can detect and eliminate all the backdoors with slight decreases in model utility.

\subsection{Effect of the Poisoning Rate}

\begin{table}[t]
    % \footnotesize
    \begin{center}
    \caption{Performance against different poisoning rates. Numbers show the relative scores compared with backdoored models.}
    \label{tab:different_poison_rates}
    % \scriptsize
    \tabcolsep=2.5pt
    \begin{tabular}{ccccc}
        \toprule
        Poisoning rates & Detected & $\Delta$ASR$\downarrow$ & $\Delta$SSIM$\downarrow$ & $\Delta$FID$\downarrow$ \\
        \midrule
        0.05 & \ding{52} & -1.00 & -1.00 & -0.07 \\
        0.10 & \ding{52} & -1.00 & -1.00 & 0.18  \\
        0.20 & \ding{52} & -1.00 & -1.00 & -0.03 \\
        0.30 & \ding{52} & -1.00 & -1.00 & 0.15  \\
        0.50 & \ding{52} & -1.00 & -1.00 & 0.20  \\
        0.70 & \ding{52} & -1.00 & -1.00 & -0.11 \\
        0.90 & \ding{52} & -1.00 & -1.00 & -0.07 \\
        % \midrule
        \bottomrule
    \end{tabular}
    \end{center}
\end{table}

We evaluate \ours{} on DMs backdoored by \baddiff{} with different poisoning rates. \cref{tab:different_poison_rates} demonstrates \ours{} can completely detect and eliminate all the backdoors and even improve the model utility in many cases.

\subsection{Effect of Clean Data}

\begin{figure}[t]
	\begin{center}
            \includegraphics[width=\linewidth]{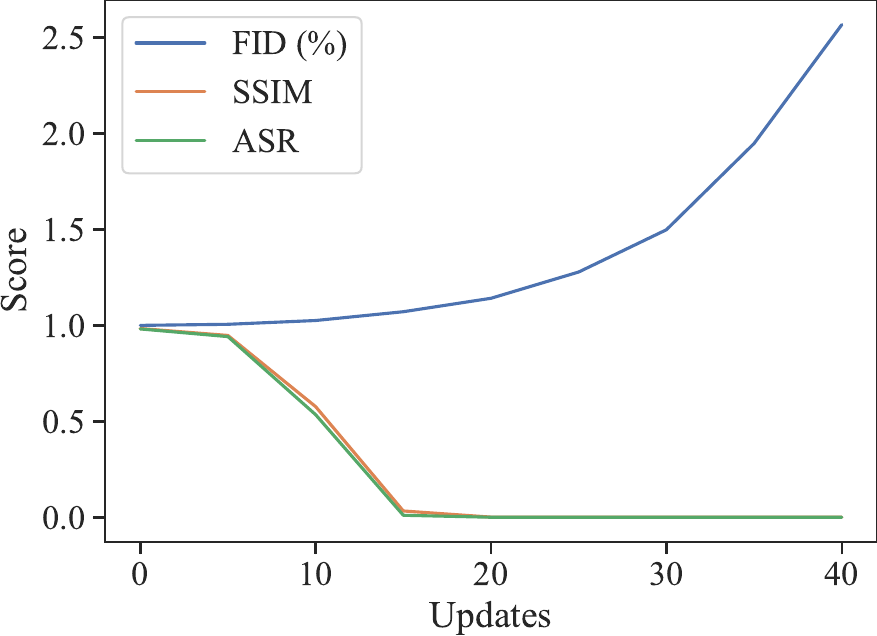}
	\end{center}
	\caption{Only use backdoor removal on a model backdoored by \baddiff{} with   stop sign trigger and   hat target.
	}
	\label{fig:only_backdoor_removal_loss}
\end{figure}

\cref{fig:only_backdoor_removal_loss} shows our backdoor removal without clean data can quickly (in 20 updates) invalidate the ground truth trigger so it cannot generate the target image. However, our trigger inversion algorithm can find another effective trigger. With the inverted trigger, the ``fixed'' model can still generate the target image.

\subsection{Effect of Backdoor Removal Loss}

\begin{figure}[t]
	\begin{center}
            \includegraphics[width=\linewidth]{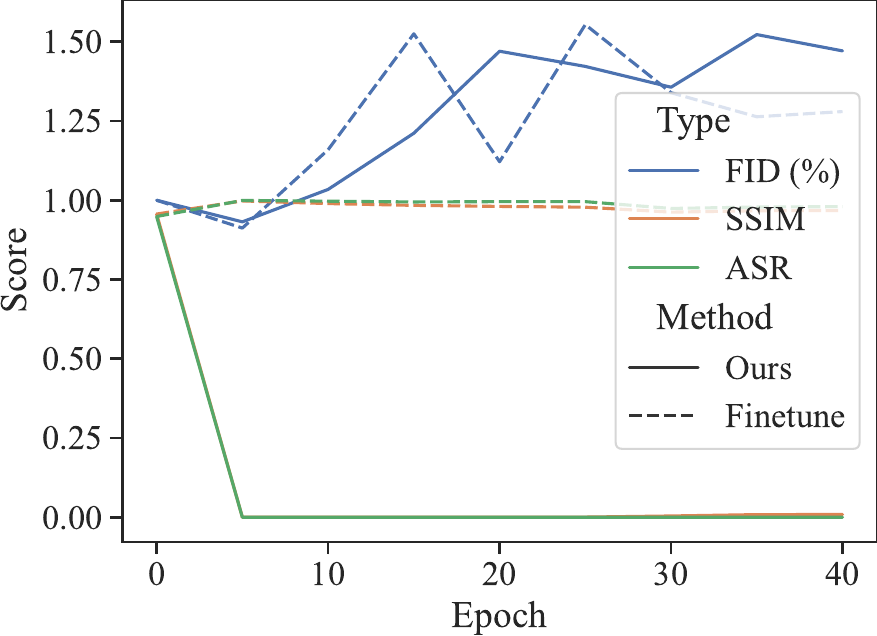}
	\end{center}
	\caption{Fine-tuning with only real data on a model backdoored by \baddiff{} with the box trigger and the hat target.
	}
	\label{fig:ours_finetune_cmp_box_hat}
\end{figure}

\begin{figure}[t]
	\begin{center}
            \includegraphics[width=\linewidth]{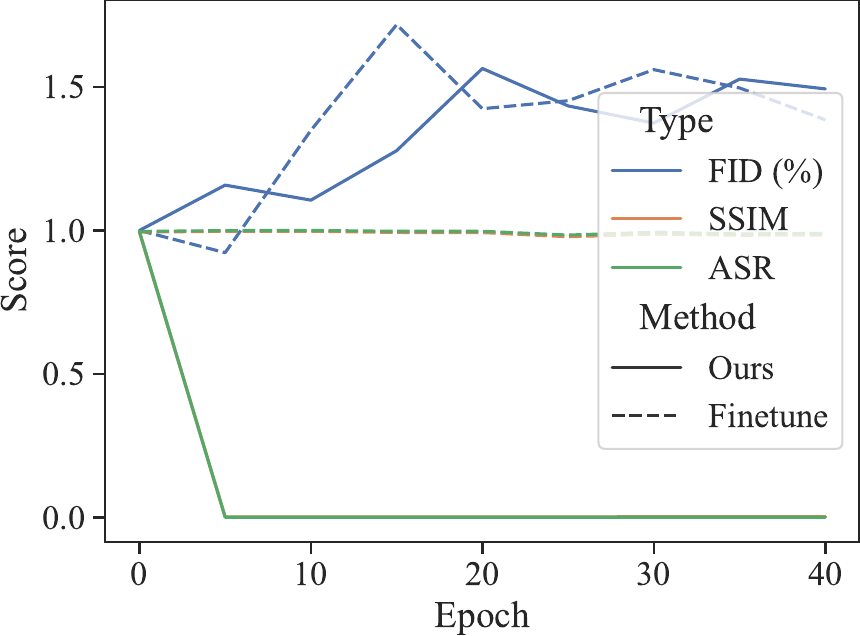}
	\end{center}
	\caption{Fine-tuning with only real data on a model backdoored by \baddiff{} with stop sign trigger and box target.
	}
	\label{fig:ours_finetune_stopsign_box}
\end{figure}

\cref{fig:ours_finetune_cmp_box_hat} and \cref{fig:ours_finetune_stopsign_box} show fine-tuning the backdoored model only with 10\% clean data cannot remove the backdoor. The green dashed line displays the ASR which is always close to 1 for the fine-tuning method, while ours (denoted by the solid green line) quickly reduces ASR to 0 within 5 epochs.

\subsection{Backdoor Removal with Real/Synthetic Data}

\begin{figure}[t]
	\begin{center}
            \includegraphics[width=\linewidth]{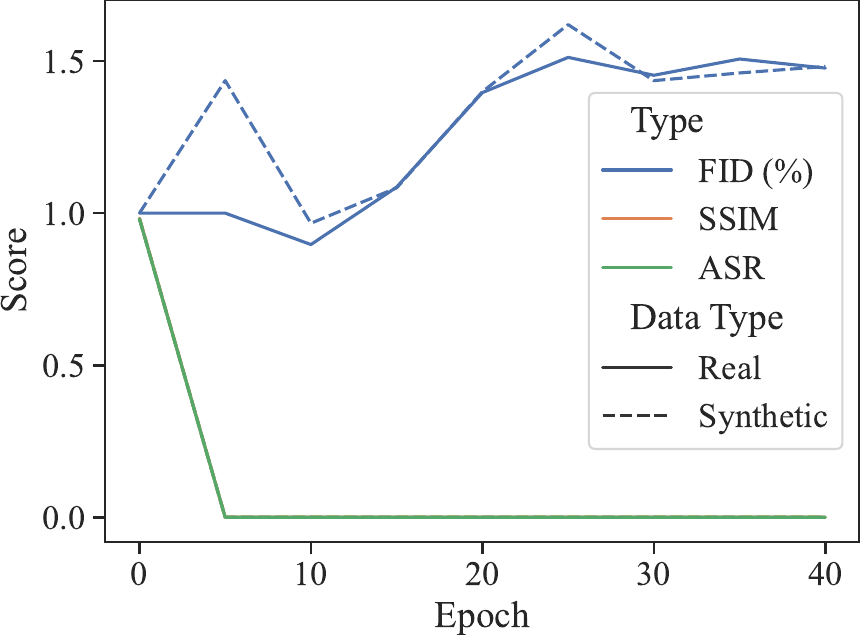}
	\end{center}
	\caption{Backdoor removal with real data or synthetic data on a model backdoored by \baddiff{} with the stop sign trigger and the hat target. ASR and SSIM lines overlap.
	}
	\label{fig:gt_synthetic_stopsign_hat}
\end{figure}

\begin{figure}[tbp]
	\begin{center}
            \includegraphics[width=\linewidth]{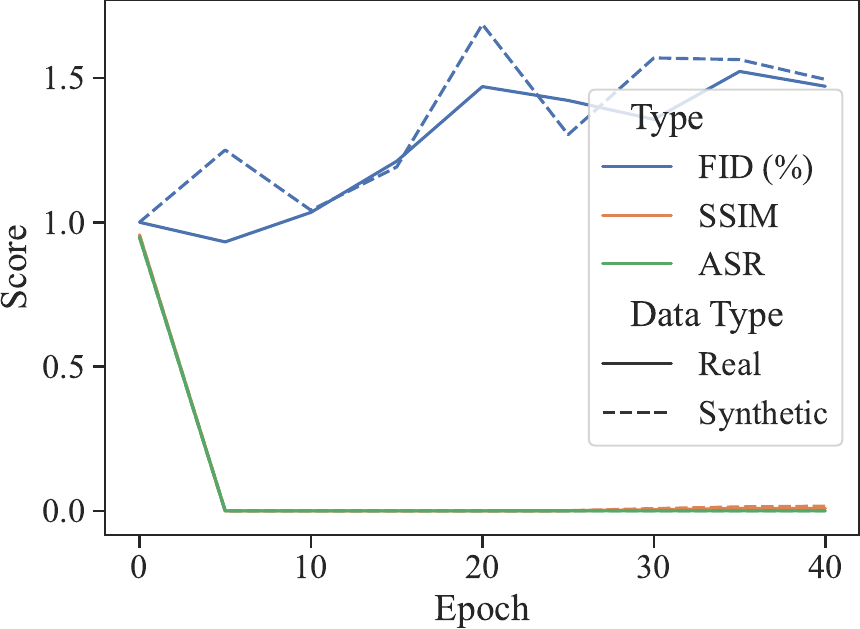}
	\end{center}
	\caption{Backdoor removal with real data or synthetic data on a model backdoored by \baddiff{} with the box trigger and the hat target. ASR and SSIM lines overlap.
	}
	\label{fig:gt_synthetic_box_hat}
\end{figure}

\cref{fig:gt_synthetic_stopsign_hat} and \cref{fig:gt_synthetic_box_hat} show more comparison between backdoor removal with real data and synthetic data. 
The overlapped SSIM and ASR lines mean the same effectiveness of backdoor removal. The FID changes in a similar trend. This means we can have a real-data-free backdoor removal approach. Since our backdoor detection is also sample-free, our whole framework can work even without access to real data.

\subsection{Detailed ODE Results}

\cref{tab:ode_results} shows the detailed results for all the ODE samplers. Our method can successfully detect the backdoored models and completely eliminate the backdoors while only slightly increasing FID.

\begin{table}[t]
    % \footnotesize
    \begin{center}
    \caption{ODE results of \villandiff{} backdoor detection and removal. Model DDIM-C means DM with DDIM sampler trained on the CIFAR-10 dataset.}
    \label{tab:ode_results}
    % \scriptsize
    \tabcolsep=2pt
    \begin{tabular}{ccccc}
        \toprule
        Model & ACC$\uparrow$ & $\Delta$ASR$\downarrow$ & $\Delta$SSIM$\downarrow$ & $\Delta$FID$\downarrow$   \\
        \midrule
DDIM-C & 1.00 & -1.00 & -1.00 & 0.14 \\
PNDM-C & 1.00 & -1.00 & -1.00 & 0.25 \\
DEIS-C & 1.00 & -1.00 & -1.00 & 0.15 \\
DPMO1-C & 1.00 & -1.00 & -1.00 & 0.15 \\
DPMO2-C & 1.00 & -1.00 & -1.00 & 0.15 \\
DPMO3-C & 1.00 & -1.00 & -1.00 & 0.15 \\
DPM++O1-C & 1.00 & -1.00 & -1.00 & 0.15 \\
DPM++O2-C & 1.00 & -1.00 & -1.00 & 0.15 \\
DPM++O3-C & 1.00 & -1.00 & -1.00 & 0.15 \\
UNIPC-C & 1.00 & -1.00 & -1.00 & 0.15 \\
HEUN-C & 1.00 & -1.00 & -1.00 & 0.25 \\
        \bottomrule
    \end{tabular}
    \end{center}
\end{table}

\subsection{Visualized results}

\begin{figure}[htbp]
	\begin{center}
            \includegraphics[width=\linewidth]{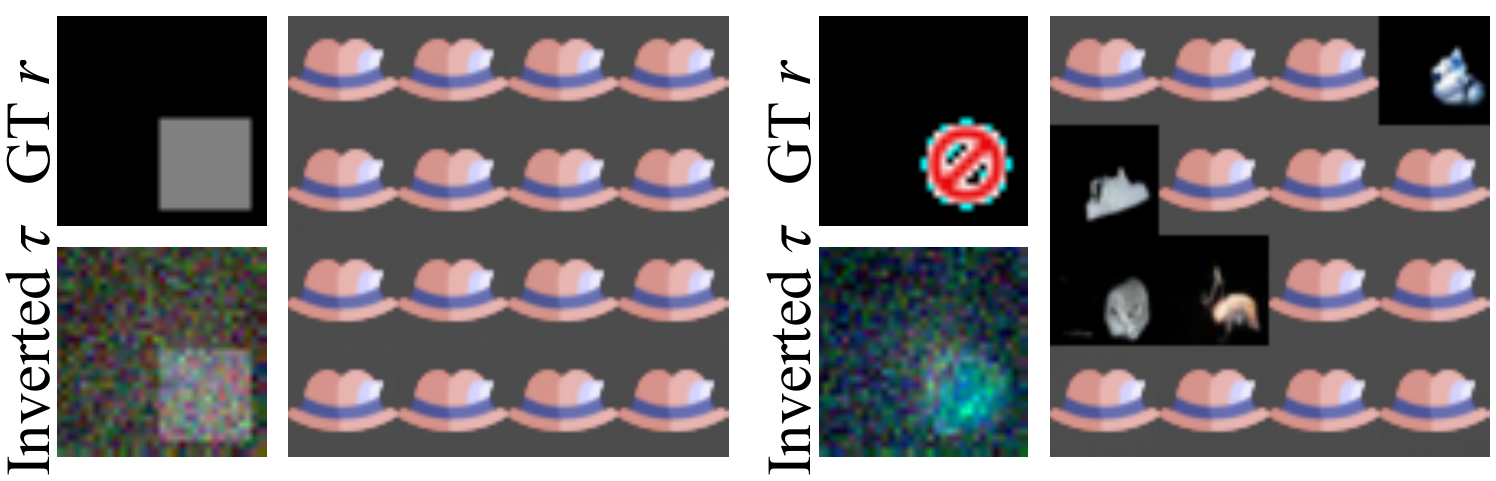}
	\end{center}
	\caption{Ground truth triggers $r$ and the corresponding inverted triggers $\tau$, as well as 16 generated images using inputs with the inverted triggers.
	}
	\label{fig:inverted_trigger}
\end{figure}

\cref{fig:inverted_trigger} visualizes some ground truth triggers and the corresponding inverted triggers. An interesting observation is that usually, the inverted trigger is not the exact same as the ground truth one. This means the injected trigger is not precise or accurate, that is, a different trigger can also trigger the backdoor effect. It's not an issue for our backdoor detection and removal framework.

\subsection{Results for Inpainting Tasks}

\baddiff{} shows they can also backdoor models used for the inpainting tasks. Given a corrupted image (\textit{e.g.}, masked with a box) without the trigger, the model can recover it into a clean image (\textit{e.g.}, complete the masked area). However, when the corrupted image contains the trigger, the target image will be generated. Our method can successfully detect the backdoored model and completely eliminate the backdoors while maintaining almost the same inpainting capability.

\subsection{Details of Adaptive Attacks}

We tried two different ways of adaptive attacks. In both cases, attackers completely know our framework. In the first case, attackers directly utilize our loss the suppress the distribution shift at the timestep $T$. Because the backdoor injection relies on the distribution shift, suppressing it will make the attack fail. In the second case, attackers choose to only inject the distribution shift starting from the timestep $T-1$ while training the timestep $T$ with only clean training loss. The intuition is our simple trigger inversion loss only uses the timestep $T$. However, this adaptive attack also failed because even if the $T-1$ step learns the distribution shift, it could not be satisfied by the $T$ step. That is, the attack also failed.

\section{Parameter Instantiation for Diffusion Models and Attacks}

\noindent
\textbf{DDPM.} This is straightforward, as DDPM is directly defined using a Markov chain. $\kappa^t = \sqrt{\alpha^t}$, $\upsilon^t=\beta^t$, $\tilde{\kappa}^t = \frac{1}{\sqrt{\alpha^t}}$, $\hat{\kappa}^t = \frac{1-\alpha^t}{\sqrt{\alpha^t} \sqrt{1-\bar{\alpha}^t}}$, $\tilde{\upsilon}^t=\frac{1-\bar{\alpha}^{t-1}}{1-\bar{\alpha}^t}\beta^t$, where $\beta^t$ is the predefined scale of noise added at step $t$, $\alpha^t = 1 - \beta^t$ and $\bar{\alpha}^t = \prod_{i=1}^t \alpha^i$. 

\noindent
\textbf{NCSN.}
$\kappa^t=1$, $\upsilon^t=(\sigma^t)^2-\sum_{i=1}^{t-1}(\upsilon^i)^2$, $\tilde{\kappa}^t=\frac{(\sigma^{t-1})^2}{(\sigma^{t-1})^2+(\upsilon^t)^2}$, $\hat{\kappa}^t = 1-\tilde{\kappa}^t$, $\tilde{\upsilon}^t = (1-\tilde{\kappa}^t)(\sigma^t)^2$, where $\sigma^t$ denotes scale of the pre-defined noise.

\noindent
\textbf{LDM.} As LDM is considered DDPM in the latent space, the instantiation is almost the same.

\noindent
\textbf{\baddiff{}}
\baddiff{} only attacks DDPM, with $\rho^t=1-\sqrt{\alpha^t}$, $\tilde{\rho}^t = \frac{(1-\sqrt{\alpha^t})\sqrt{1-\bar{\alpha}^t}}{\alpha^t-1}$.

\noindent
\textbf{\trojdiff{}}
$\rho^t=k^t$, $\upsilon^t_b=\beta^t\gamma^2$, $\tilde{\kappa}^t_b=\frac{\sqrt{\alpha^t}(1-\bar{\alpha}^{t-1})}{1-\bar{\alpha}^t}+\frac{1}{\sqrt{\bar{\alpha}^t}}$, $\hat{\kappa}^t_b=\frac{-\sqrt{1-\bar{\alpha}^t} (\gamma}{\sqrt{\bar{\alpha}^t}}$, $\tilde{\rho}^t = \frac{\sqrt{1-\bar{\alpha}^t}}{\sqrt{\bar{\alpha}^t}} + \frac{\sqrt{1-\bar{\alpha}^{t-1}}\beta^t - \sqrt{\alpha^t}(1-\bar{\alpha}^{t-1}k^t)}{1-\bar{\alpha}^t}$, where $k^t = \sqrt{1 - \bar{\alpha}^t} - \sum_{i=2}^t \prod_{j=i}^t\sqrt{\alpha^j}k^{i-1}$. Note the subscript shows a different chain for the backdoor from the clean chain.

\noindent
\textbf{\villandiff{}}
It considers a more general set of DMs and assumes the trigger distribution shift is $\hat{\rho}^t r$. That is, the clean forward chain is $q(x_c^t|x_c^0) = \mathcal{N}(x_c^t; \hat{\alpha}^t x^0_c, \hat{\beta}^t I)$ and the backdoor forward chain is $q(x_b^t|x_b^0) = \mathcal{N}(x_b^t; \hat{\alpha}^t x^0_c + \hat{\rho}^t r,  \hat{\beta}^t I)$. Then $\kappa^t=\frac{\hat{\alpha}^t}{\hat{\alpha}^{t-1}}$, $\upsilon^t=\hat{\beta}^t-\sum_{i=1}^{t-1}((\prod_{j=i+1}{t}\kappa^j)^2\upsilon^i)$, $\tilde{\kappa}^t = \frac{\kappa^t\hat{\beta}^{t-1}}{(\kappa^t)^2\hat{\beta}^{t-1}+\upsilon^t}$, $\hat{\kappa}^t=\frac{\hat{\alpha}^{t-1}\upsilon^t}{(\kappa^t)^2\hat{\beta}^{t-1}+\upsilon^t}$, $\tilde{\upsilon}^t=\frac{\hat{\kappa}^t}{\hat{\alpha}^t}\hat{\beta}^t$.

\section{Summary of used Symbols}

% \begin{table}[htbp]
\begin{center}
\small
% \caption{Meaning of Symbols}
% \label{tab:symbols}
% \vspace{-10pt}
\tabcolsep=2pt
\begin{tabular}{r|l}
\toprule
% Symbol & Meaning \\
% \midrule
$\mathcal{N}(\mu, \sigma)$ & Normal distribution with mean $\mu$ and std $\sigma$ \\
\midrule
$x^t$, $x^t_{c}$, $x^t_{b}$ & DM's output at step $t$, $c$/$b$ means clean/backdoor  \\ 
\midrule 
$M_{\theta}$ & UNet $M$ with parameter $\theta$ (omitted sometimes)  \\
\midrule
$q(x^t|x^{t-1})$ & Forward probability defined by DM \\
\midrule
$p_{\theta}(x^{t-1}|x^t)$ & Reverse probability based on $M_{\theta}$ \\
\midrule
$r$ or $\tau$ & Ground truth injected or inverted trigger \\
\midrule
$\lambda$ & The linear dependence coefficient \\
\midrule
$x_{[1, n]}$ & A set (or a tensor) of $n$ $x$'s \\ 
\midrule
$\kappa^t$, $\hat{\kappa}^t$, $\tilde{\kappa}^t$ & Transitional content schedulers in DM \\
\midrule
$\upsilon^t$, $\tilde{\upsilon}^t$ & Transitional noise schedulers in DM \\
\midrule
$\rho^t$, $\tilde{\rho}^t$ & Scale of distribution shift \\
\bottomrule
\end{tabular}
\end{center}
% \end{table}

\section{Limitation of Existing Methods on DMs}

Here we use the simplified NC\footnote{https://github.com/bolunwang/backdoor} on a cat-dog classifier $f$ as an example. NC first generates a trigger for each label. E.g., $\tau_{\textcolor{red}{\text{cat}}}=\arg \min_{\tau} \sum_{x\in \textcolor{red}{X^{\text{dog}}_{[1, n]}}} \ell (\textcolor{red}{\text{cat}}, f(x+\tau))$, i.e., when $\tau_{\text{cat}}$ is added to $n$ dog's images, $f$ misclassifies them as cat\footnote{$\ell$ is the classification loss such as cross-entropy.}. 
NC regards the model as backdoored if there's a significantly small trigger for one label, \textit{e.g.} $\tau_{\text{cat}} \ll \tau_{\text{dog}}$.
NC needs the label information and clean samples (highlighted in red) while DMs don't output labels or have clean samples belonging to a victim class.
%\ours{} doesn't require clean samples, so 
It hence can't be easily adapted to DMs.

\end{document}